\documentclass[12pt,a4paper,numbers]{elsarticle}
\biboptions{sort&compress} 

\usepackage{amsmath,amssymb,amsfonts}
\usepackage{bm}
\usepackage{graphicx}
\usepackage{subcaption}
\usepackage{booktabs}
\usepackage{xcolor}
\usepackage[colorlinks=true, linkcolor=black, citecolor=black, urlcolor=blue, breaklinks=true]{hyperref}
\usepackage{times}    
\usepackage{doi}       
\usepackage{tabularx}
\usepackage[normalem]{ulem}

\newcommand{\etac}{\eta_{\text{Carnot}}}
\newcommand{\epsl}{\varepsilon}
\newcommand{\etab}{\xi}
\newcommand{\paramu}{u}
\newcommand{\paramv}{v}

\begin{document}

\begin{frontmatter}

\title{Quantum Otto engine powered by an anisotropic Heisenberg XYZ model under independent local magnetic fields}

\author[1]{Meiru Li}
\author[1]{Maimaitiyiming Tusun\corref{cor1}}
\author[1]{Fang Zhao\corref{cor2}}
\author[2]{Hasiyatihan Abudoula}
\author[1]{Tongcheng Wei}

\cortext[cor1]{Corresponding author. Email: hawk@xjnu.edu.cn.}
\cortext[cor2]{Corresponding author. Email: gluons@foxmail.com.} 

\address[1]{School of Physics and Electronic Engineering, Xinjiang Normal University, Xinjiang Key Laboratory of Luminescent Minerals and Optical Functional Materials, Urumqi 830054, China}
\address[2]{Urumqi No. 78 Middle School, Urumqi 830054, China}

\begin{abstract}
We study a quantum Otto heat engine whose working substance is an anisotropic two-qubit Heisenberg XYZ model. Independent local magnetic fields are used to control each spin individually. The influence of the longitudinal coupling, anisotropy, transverse coupling, and local fields on the net work output and efficiency is systematically examined. Reducing the longitudinal coupling is found to markedly improve both the maximum work and the peak efficiency. The engine performance reaches an optimum at a particular value of the anisotropy parameter. A local work analysis clarifies how work is produced during the cycle. Because of the asymmetric local fields and the intrinsic spin-spin interaction, the two qubits play markedly different thermodynamic roles; the interaction term itself contributes crucially to the total work. We further analyze the variation of quantum entanglement, quantified by concurrence, along the cycle. The results indicate that a pronounced change in entanglement between the hot and cold isomagnetic strokes is closely correlated with the efficiency enhancement. This work offers new insight into the operating principles and control of quantum Otto heat engines.
\end{abstract}


\end{frontmatter}

\section{Introduction}

Quantum thermodynamics (QT) studies thermodynamic processes within the framework of quantum dynamics \cite{Kosloff2013, Cangemi2024, Quan2023, Gemmer2009, Vinjanampathy2016, Binder2018}. A quantum heat engine (QHE) is a central platform for exploring the foundations of QT, as it extends classical thermodynamic cycles to the microscopic domain. Several quantum thermodynamic cycles have been proposed and demonstrated experimentally \cite{Uzdin2015, Scully2003, Kieu2004}. Among them, the Otto cycle has become a paradigmatic model owing to its conceptual clarity and analytical tractability \cite{AbdRabbou2023, Quan2007, Quan2009, Kieu2004, Zhang2007, Feldmann2000, Feldmann2003}. Many quantum systems have been employed as working substances for work extraction, including pairs of coupled spins \cite{Feldmann2004}, single spins \cite{Ono2020}, harmonic oscillators \cite{Wang2007, Wu2009}, ensembles of nitrogen-vacancy centers in diamond \cite{Klatzow2019}, and Heisenberg spin models \cite{He2012, Huang2018, Thomas2011}. Over the past decade, quantum dots \cite{Jaegon2022}, superconducting circuits \cite{Uusnakki2025a, Vadimov2025, Rasola2025, Uusnakki2026, Sundelin2026}, and Heisenberg XX and XXZ models with Dzyaloshinskii-Moriya interaction (DMI) have also attracted considerable attention \cite{Dzyaloshinskii1958, Moriya1960, Peng2019, Ahadpour2021, Kuznetsova2023, Zhao2017, Zhang2008, Khlifi2020}. Beyond cyclic engines, Bresque \textit{et al.} \cite{Bresque2025} recently investigated a lazy quantum measurement engine theoretically, deriving exact work statistics and revealing run-and-tumble behaviors that illuminate the thermodynamic cost of information processing.

The two- and three-dimensional Heisenberg models have been examined in many contexts. For example, the influence of magnetic fields on entanglement in the Heisenberg model was studied in Refs. \cite{Kamta2002, Arnesen2001}, and the influence of magnetic fields, with and without the Dzyaloshinskii-Moriya interaction, on thermal entanglement has been analyzed \cite{Li2008, Abliz2006, Xu2014, Sun2003, Zhou2003}. Spin-orbit couplings, such as the Dzyaloshinskii-Moriya (DM) and Kaplan-Shekhtman-Entin-Wohlman-Aharonov (KSEA) interactions, also play an important role in shaping the thermodynamic landscape. Recent analyses of Stirling engines have shown that these interactions, together with the Heisenberg exchange coupling $J$, govern the transitions among distinct operational modes (heat engine, refrigerator, or heater) \cite{Xiao2026, Tang2026}. Altintas and M\"{u}stecapl{\i}o\u{g}lu \cite{Altintas2015} established a general local-thermodynamics formalism and applied it to a quantum Otto engine with a spin-1/2 coupled to an arbitrary spin, while Kuznetsova \textit{et al.} \cite{Kuznetsova2023} examined an Otto QHE with an XYZ working medium that includes DMI and KSEA interactions. Makouri \textit{et al.} \cite{ElMakouri2023} showed that the performance of a coupled Otto QHE can be enhanced even when entanglement and quantum correlations are absent. More recently, Tarif \textit{et al.} \cite{Tarif2025, Tarif2026} have examined two-qubit thermal machines from complementary perspectives: an Otto engine whose operational modes are controlled by coherence and entanglement \cite{Tarif2025}, and a two-qubit system coupled to a single-mode cavity with Stark shifts and Dzyaloshinskii-Moriya interaction, in which quantum coherence, quantum discord, and heat transport are characterized jointly within a quantum Otto cycle \cite{Tarif2026}. These studies underline the central role of quantum correlations in few-qubit machines and motivate the quantitative entanglement-performance analysis presented below.

The properties of QHEs have been explored extensively for different working substances and environmental couplings. Khlifi \textit{et al.} \cite{Khlifi2020} considered a QHE based on the XXX, XXZ, and XYZ Heisenberg models with two particles, and demonstrated that both DMI and the magnetic field affect the efficiency and work output. Huang \textit{et al.} \cite{Huang2018} examined an Otto QHE with a three-qubit Heisenberg XXZ model with DMI in a magnetic field. Ahadpour \textit{et al.} \cite{Ahadpour2021} investigated a four-level entangled quantum refrigerator within the XYZ Heisenberg models. Beyond the intrinsic Hamiltonian parameters, engineering the system-environment interaction and the configuration of external fields has emerged as a powerful control strategy. For example, coupling the working substance to non-equilibrium squeezed thermal reservoirs can create resonance windows that significantly boost work extraction \cite{Zhu2026}. Moreover, inhomogeneous magnetic fields, analyzed within both Shannon and Tsallis entropy frameworks, have been found to induce spin localization and suppress global coherence, thereby altering the heat-work balance \cite{Han2026}. In the specific context of Otto engines, the distinction between local and common reservoir couplings has also proved decisive in determining whether the machine operates as an engine or a heater, with the orientation of the interaction axis acting as a strong amplification mechanism \cite{Mashhor2025}.

Despite these advances, the interplay between local work distribution and quantum correlations in an Otto cycle with an anisotropic XYZ working substance under independent local magnetic fields has not been systematically explored. In particular, the extent to which entanglement can serve as an indicator, or witness, of the energy-level restructuring that governs work extraction remains an open question. We study this system for several reasons. (i) Although the quantum Otto cycle is one of the most studied cycles, the effect of independent local magnetic fields on its performance is still poorly understood. (ii) The anisotropic XYZ model offers a rich parameter space that allows precise control over the energy-level structure. (iii) The exact solvability of the two-qubit XYZ model enables us to compute both thermodynamic quantities and quantum entanglement analytically, providing a unique opportunity to examine their relationship without numerical ambiguity.

The main contributions of this work are threefold. (i) Using the exact eigenstates of the two-qubit anisotropic XYZ model with independent local fields, we systematically analyze how $J_z$, $\Delta$, $J$, and $(B_1,B_2)$ influence the net work output $W_{\text{net}}$ and efficiency $\eta$, and we identify the optimal operating regions. (ii) We carry out a comprehensive local work analysis that reveals how energy is partitioned between the two qubits and their interaction, uncovering a robust asymmetric differentiation of their thermodynamic roles that is linked to the local field asymmetry. (iii) By evaluating the concurrence along the Otto cycle and deriving its simplified analytical form for the X-state thermal density matrix, we investigate the correlation between entanglement dynamics and engine performance, and we clarify that the entanglement asymmetry acts as a witness of the underlying energy-level restructuring rather than as an independent thermodynamic resource. These results provide practical guidance for optimizing solid-state quantum heat engines and deepen the understanding of quantum correlations in thermodynamic cycles.

\section{Theory and model}
\label{sec: theory and model}

In natural units ($\hbar=1$), the Hamiltonian of the anisotropic two-qubit Heisenberg XYZ model reads
\begin{equation}\label{eq:H_original}
    \hat{H} = J_x\hat{S}_1^x\hat{S}_2^x + J_y\hat{S}_1^y\hat{S}_2^y + J_z\hat{S}_1^z\hat{S}_2^z + B_1\hat{S}_1^z + B_2\hat{S}_2^z,
\end{equation}
where $\hat{S}_i^\alpha = \frac{1}{2}\hat{\sigma}_i^\alpha$ and $\hat{\sigma}_i^\alpha$ are the Pauli matrices ($\alpha=x,y,z$). The coupling parameters $J_x$, $J_y$, $J_z$ represent the spin-exchange interaction strengths along the $x$, $y$, and $z$ directions, respectively, while $B_1$ and $B_2$ are independent local magnetic fields acting on the two spins.

With the substitutions
\[
J = \frac{J_x + J_y}{2},\quad \Delta = \frac{J_x - J_y}{2},\quad \hat{S}_i^{\pm} = \hat{S}_i^x\pm i\hat{S}_i^y,
\]
where $J$ characterizes the average in-plane coupling strength and $\Delta$ measures the degree of in-plane anisotropy, the Hamiltonian takes the alternative form
\begin{equation}\label{eq:hamiltonian}
\begin{aligned}
\hat{H} &= \frac{J}{2}\left(\hat{S}_1^+\hat{S}_2^- + \hat{S}_1^-\hat{S}_2^+\right) + \frac{\Delta}{2}\left(\hat{S}_1^+\hat{S}_2^+ + \hat{S}_1^-\hat{S}_2^-\right) \\
&+ J_z\hat{S}_1^z\hat{S}_2^z + B_1\hat{S}_1^z + B_2\hat{S}_2^z,
\end{aligned}
\end{equation}
which is convenient for the subsequent physical analysis. The nonlinear interactions present in this model are analogous to those encountered in superconducting quantum circuits; recent advances in nonlinear-response theory have provided powerful tools for analyzing such dissipative and driven quantum systems \cite{Vadimov2025}.

Table~\ref{tab:parameters} lists the model parameters, their physical meaning, and the typical ranges employed in this work.

\begin{table}[htbp]
\centering
\caption{Model parameters: definitions, physical interpretations, and typical numerical ranges. Negative values of $B_{1,2}$ correspond to reversed local field directions.}
\label{tab:parameters}
\begin{tabular}{cclc}
\toprule
Symbol & Definition & Physical meaning & Typical range \\
\midrule
$J$ & $(J_x+J_y)/2$ & Average $XY$ coupling strength & $[0, 8.0]$ \\
$\Delta$ & $(J_x-J_y)/2$ & $XY$ anisotropy parameter & $[0, 5.0]$ \\
$J_z$ & $J_z$ & $z$-direction coupling strength & $[0.5, 1.0]$ \\
$B_1$, $B_2$ & local fields & Independent fields on two spins & $[-1.0, 4.0]$ \\
$T_h$, $T_c$ & bath temperatures & Hot and cold reservoir temperatures & $T_h = 2.0$, $T_c = 1.0$ \\
\bottomrule
\end{tabular}
\end{table}

It is important to assess the experimental feasibility of the parameter ranges used here; we adopt superconducting circuits as the reference platform, since every term of Eq.~(\ref{eq:hamiltonian}) is accessible with independently controlled parameters. A pair of transmons coupled through a tunable coupler provides (i) the local fields $B_1$ and $B_2$ as the flux-bias detunings of the two qubits, set independently by two dedicated bias lines --- values in the range $0.2$--$0.75$ GHz, including zero and sign reversal, are routine \cite{Yan2018}; (ii) the exchange $J$ as a parametrically activated flip--flop interaction \cite{Caldwell2018}; (iii) the anisotropy $\Delta$ as the parametrically activated double-excitation (two-photon) term $\frac{\Delta}{2}(\hat{S}_1^+\hat{S}_2^+ + \hat{S}_1^-\hat{S}_2^-)$ at the qubit sum frequency \cite{Roth2017}; and (iv) the Ising coupling $J_z$ through the static $ZZ$ interaction of the coupler, tunable in the required sub-dominant range $J_z < J$ \cite{Sung2021}. Taking the energy unit as $h\times0.25$ GHz, the working temperatures correspond to $T_c = 1 \simeq 12$ mK and $T_h = 2 \simeq 24$ mK, the regime in which engineered thermal reservoirs for superconducting qubits have been demonstrated \cite{Rasola2025, Uusnakki2026, Sundelin2026}. Dipolarly coupled spin systems such as NV centers cannot realize the anisotropy term (dipolar exchange is rotationally symmetric in the transverse plane, $J_x = J_y$), share a common global field, and do not provide thermal reservoirs; they are therefore not pursued here. Negative local magnetic fields (e.g., $B_{2h} < 0$), which appear in our parameter scans, do not represent unphysical conditions; they simply correspond to a reversal of the local effective field direction, which can be realized by applying a $\pi$-phase shift to the local microwave driving field or by using opposite external bias currents in flux-tunable superconducting circuits \cite{Rasola2025}.

Introducing the sum and difference fields $\epsl = B_1+B_2$ and $\etab = B_1-B_2$, and working in the computational basis $\{|00\rangle, |01\rangle, |10\rangle, |11\rangle\}$, the Hamiltonian can be written in block-diagonal matrix form as
\begin{equation}\label{eq:H_matrix}
\hat{H} = \begin{pmatrix}
\frac{J_z}{4} + \frac{\epsl}{2} & 0 & 0 & \frac{\Delta}{2} \\[4pt]
0 & -\frac{J_z}{4} + \frac{\etab}{2} & \frac{J}{2} & 0 \\[4pt]
0 & \frac{J}{2} & -\frac{J_z}{4} - \frac{\etab}{2} & 0 \\[4pt]
\frac{\Delta}{2} & 0 & 0 & \frac{J_z}{4} - \frac{\epsl}{2}
\end{pmatrix}.
\end{equation}
The diagonal elements $-\frac{J_z}{4} \pm \frac{\etab}{2}$ belong to the subspace spanned by $\{|01\rangle, |10\rangle\}$; here the opposite spin projections give the $-J_z/4$ contribution from the Ising interaction, while the $\pm\etab/2$ terms arise from the local field difference acting on the respective single-spin states. Diagonalizing this matrix yields four eigenvalues,
\begin{align}
E_1 &= -\frac{J_z}{4} - \frac{\paramv}{2}, \quad 
E_2 = -\frac{J_z}{4} + \frac{\paramv}{2}, \label{eq:eig1}\\
E_3 &= \frac{J_z}{4} - \frac{\paramu}{2}, \quad 
E_4 = \frac{J_z}{4} + \frac{\paramu}{2}, \label{eq:eig2}
\end{align}
with $\paramu = \sqrt{\epsl^2 + \Delta^2}$ and $\paramv = \sqrt{\etab^2 + J^2}$. The corresponding normalized eigenstates can be expressed compactly as
\begin{align}
|\psi_1\rangle &= \frac{1}{\sqrt{J^2 + (\etab - \paramv)^2}}\begin{pmatrix} 0 \\ \etab - \paramv \\ J \\ 0 \end{pmatrix}, \label{eq:psi1}\\
|\psi_2\rangle &= \frac{1}{\sqrt{J^2 + (\etab + \paramv)^2}}\begin{pmatrix} 0 \\ \etab + \paramv \\ J \\ 0 \end{pmatrix}, \label{eq:psi2}\\
|\psi_3\rangle &= \frac{1}{\sqrt{\Delta^2 + (\epsl - \paramu)^2}}\begin{pmatrix} \epsl - \paramu \\ 0 \\ 0 \\ \Delta \end{pmatrix}, \label{eq:psi3}\\
|\psi_4\rangle &= \frac{1}{\sqrt{\Delta^2 + (\epsl + \paramu)^2}}\begin{pmatrix} \epsl + \paramu \\ 0 \\ 0 \\ \Delta \end{pmatrix}. \label{eq:psi4}
\end{align}
We emphasize that the normalization coefficients of these eigenstates have been rigorously corrected relative to the expressions given in Ref. \cite{Abliz2006}; this correction guarantees the exact orthonormality required for accurate thermodynamic evaluations. The labeling of the eigenstates follows the energy ordering $E_1 < E_3 < E_2 < E_4$ for the typical parameter regime ($J>0, J_z>0, B_1,B_2>0$).

At thermal equilibrium with a bath at temperature $T=1/(k_B\beta)$, the partition function is $Z = \sum_{i=1}^4 e^{-\beta E_i}$, and the occupation probability of each eigenstate is $p_i = e^{-\beta E_i}/Z$. The internal energy then follows as
\begin{equation}
U = \sum_{i=1}^4 p_i E_i.
\end{equation}
For a quantum system, the first law of thermodynamics takes the form $dU = \delta Q + \delta W = \sum_i E_i dp_i + \sum_i p_i dE_i$, where $\delta Q = \sum_i E_i dp_i$ is the absorbed heat and $\delta W = \sum_i p_i dE_i$ is the work done on the system \cite{Quan2023}. 

The partition function can be written in the compact analytical form
\begin{equation}
Z = 2e^{-\beta J_z/4}\cosh\left(\frac{\beta \paramu}{2}\right) + 2e^{\beta J_z/4}\cosh\left(\frac{\beta \paramv}{2}\right). \label{eq:partition}
\end{equation}
Furthermore, the thermal density matrix $\hat{\rho}(T) = e^{-\beta \hat{H}}/Z$ can be obtained in closed form without summing explicitly over eigenstates. In the computational basis it remains block-diagonal:
\begin{equation}
\hat{\rho}(T) = \begin{pmatrix}
\rho_{11} & 0 & 0 & \rho_{14} \\
0 & \rho_{22} & \rho_{23} & 0 \\
0 & \rho_{32} & \rho_{33} & 0 \\
\rho_{41} & 0 & 0 & \rho_{44}
\end{pmatrix},
\end{equation}
with the non-zero matrix elements given by
\begin{align}
\rho_{11} &= \frac{e^{-\beta J_z/4}}{Z} \left[ \cosh\left(\frac{\beta \paramu}{2}\right) - \frac{\epsl}{\paramu} \sinh\left(\frac{\beta \paramu}{2}\right) \right], \label{eq:rho11}\\
\rho_{44} &= \frac{e^{-\beta J_z/4}}{Z} \left[ \cosh\left(\frac{\beta \paramu}{2}\right) + \frac{\epsl}{\paramu} \sinh\left(\frac{\beta \paramu}{2}\right) \right], \label{eq:rho44}\\
\rho_{14} &= \rho_{41}^* = -\frac{e^{-\beta J_z/4}}{Z} \cdot \frac{\Delta}{\paramu} \sinh\left(\frac{\beta \paramu}{2}\right), \label{eq:rho14}\\
\rho_{22} &= \frac{e^{\beta J_z/4}}{Z} \left[ \cosh\left(\frac{\beta \paramv}{2}\right) - \frac{\etab}{\paramv} \sinh\left(\frac{\beta \paramv}{2}\right) \right], \label{eq:rho22}\\
\rho_{33} &= \frac{e^{\beta J_z/4}}{Z} \left[ \cosh\left(\frac{\beta \paramv}{2}\right) + \frac{\etab}{\paramv} \sinh\left(\frac{\beta \paramv}{2}\right) \right], \label{eq:rho33}\\
\rho_{23} &= \rho_{32}^* = -\frac{e^{\beta J_z/4}}{Z} \cdot \frac{J}{\paramv} \sinh\left(\frac{\beta \paramv}{2}\right). \label{eq:rho23}
\end{align}
For real parameters $J, \Delta, J_z, B_1, B_2$, all matrix elements are real, so $\rho_{41}=\rho_{14}$ and $\rho_{32}=\rho_{23}$. The derivation of these expressions is outlined in Appendix~\ref{app:rho_derivation}.

\section{Quantum Otto cycle}
The quantum Otto cycle is a canonical model for studying work--heat conversion in the quantum regime. It consists of two adiabatic strokes, during which the system is isolated and the external parameters are varied slowly enough that the quantum adiabatic theorem holds---no transitions occur between instantaneous eigenstates---and two isomagnetic strokes, during which the magnetic fields are kept fixed and the system thermalizes with a reservoir \cite{Quan2007, Quan2009, Cangemi2024, Kuznetsova2023}. 

We focus here on the ideal adiabatic limit, which provides the necessary theoretical benchmark for identifying the fundamental mechanisms of energy partitioning. In any practical finite-time implementation, non-adiabatic transitions---often referred to as quantum friction---will inevitably generate irreversible entropy and consume some of the extractable work \cite{Feldmann2000, Feldmann2003}. Such quantum friction would appear as an additional dissipation channel that reduces the magnitude of the local work components $w_1$ and $w_2$, and could blur the sharp functional differentiation between the two qubits observed in the ideal cycle. Nevertheless, the qualitative picture of asymmetric energy partitioning driven by the local field imbalance is expected to persist provided the driving remains adiabatic. For the linear ramps $B_i(t)=B_{ic}+\frac{t}{\tau}(B_{ih}-B_{ic})$, the adiabatic condition $\max_{m\neq n}|\langle m|\partial_t\hat{H}|n\rangle|/\Delta E_{mn}^2 \ll 1$, together with $\partial_{B_i}\hat{H}=\hat{S}_i^z$ bounded by $1/2$, reduces to the explicit requirement
\[
\tau \gg \frac{|\Delta B_1| + |\Delta B_2|}{2\,\Delta E_{\min}^2},
\]
where $\Delta E_{\min}$ is the smallest instantaneous gap along the stroke. The dotted segments in Fig.~\ref{fig:eta_vs_J_Delta} mark the ranges of $J$ in which the required driving time exceeds a practical bound $\tau_{\min}\simeq 10^{3}$, i.e., where $\Delta E_{\min}$ falls below $\simeq 0.05$ in the dimensionless units of the figure; this includes the subranges in which the gap closes during the stroke ($\tau\to\infty$). Outside these segments the above requirement is readily met. A quantitative treatment of finite-time effects is deferred to future work.

Although most studies of quantum Otto engines rely on external periodic driving, recent work has demonstrated that autonomous operation is possible by exploiting nonlinear couplings between the working substance and a flywheel oscillator, as realized in the superconducting circuit experiment of Ref. \cite{Uusnakki2026}. In the present paper we concentrate on the standard externally driven Otto cycle, which offers a well-controlled framework for studying how local magnetic field tuning affects engine performance. For the two-qubit Heisenberg XYZ system under study---where the external fields $B_1$ and $B_2$ serve as the control parameters---a complete Otto cycle consists of the four steps sketched in Fig.~\ref{fig:cycle}. The physical interpretation of each step is given below.

\begin{figure}[htbp]
\centering
\includegraphics[width=0.9\linewidth]{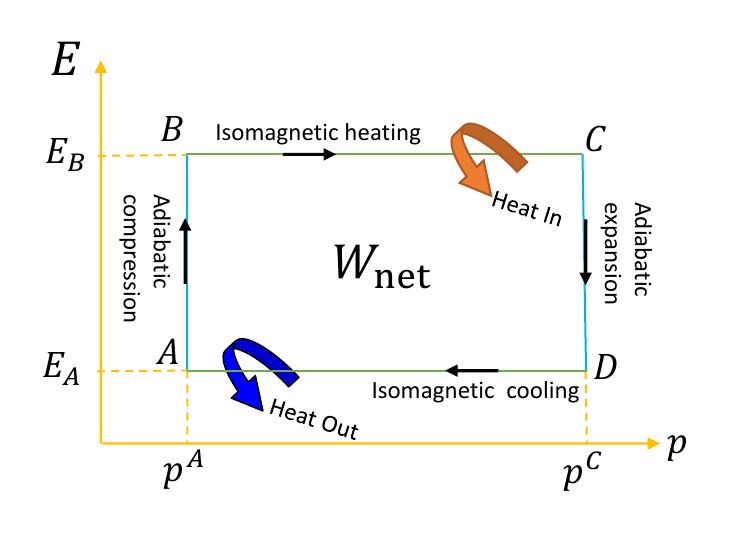}
\caption{Sketch of the quantum Otto cycle.}
\label{fig:cycle}
\end{figure}

\begin{enumerate}
\item \textbf{Adiabatic compression} ($A \to B$): The system, initially in equilibrium with the cold reservoir at temperature $T_c$ (state $A$), is isolated. By the quantum adiabatic theorem, the occupation probabilities $p_i$ of the instantaneous eigenstates remain unchanged during the adiabatic evolution. The magnetic fields are varied from $(B_{1c},B_{2c})$ to $(B_{1h},B_{2h})$ (with $\epsl_h > \epsl_c$), thereby enlarging the level spacings. The eigenenergies change as
\[
E_i^A = E_i(B_{1c},B_{2c}), \qquad E_i^B = E_i(B_{1h},B_{2h}).
\]
The populations at stage $A$ are given by the thermal distribution
\begin{equation}
p_i^A = \frac{e^{-\beta_c E_i^A}}{Z_c}, \qquad Z_c = \sum_{i=1}^4 e^{-\beta_c E_i^A}. \label{eq:pA}
\end{equation}
Because the process is adiabatic, $p_i^B = p_i^A$. The work done on the system is
\[
W_{AB} = \sum_{i=1}^4 p_i^A (E_i^B - E_i^A).
\]

\item \textbf{Isomagnetic heating} ($B \to C$): The system is brought into contact with the hot reservoir at temperature $T_h$ while the fields remain fixed at $(B_{1h},B_{2h})$. Thermalization at constant energy levels drives the system into the Gibbs state $C$ with probabilities
\begin{equation}
p_i^C = \frac{e^{-\beta_h E_i^B}}{Z_h}, \qquad Z_h = \sum_{i=1}^4 e^{-\beta_h E_i^B}. \label{eq:pC}
\end{equation}
No work is performed because the level structure is unchanged; the system absorbs heat from the hot bath,
\[
Q_h = \sum_{i=1}^4 E_i^B (p_i^C - p_i^B).
\]

\item \textbf{Adiabatic expansion} ($C \to D$): The system is isolated again and the magnetic fields are adiabatically returned to their original values $(B_{1c},B_{2c})$. The level spacings revert to their initial sizes and the populations remain unchanged, $p_i^D = p_i^C$. The system delivers work to the external agent:
\[
W_{CD} = \sum_{i=1}^4 p_i^C (E_i^A - E_i^B).
\]

\item \textbf{Isomagnetic cooling} ($D \to A$): Finally, the system is coupled to the cold reservoir at temperature $T_c$ with the fields fixed at $(B_{1c},B_{2c})$. It thermalizes back to the initial equilibrium state $A$, releasing heat to the cold bath:
\[
Q_c = \sum_{i=1}^4 E_i^A (p_i^A - p_i^D).
\]
\end{enumerate}

We note that only the states $A$ and $C$ are Gibbs states described by the thermal density matrix of Eqs.~(\ref{eq:rho11})--(\ref{eq:rho23}). The states $B$ and $D$ are generally not: since the adiabatic strokes preserve the occupation probabilities in the instantaneous eigenbasis, their density operators are
\begin{equation}
\hat{\rho}_B = \sum_{i=1}^{4} p_i^A\,|\psi_i^B\rangle\langle\psi_i^B|, \qquad
\hat{\rho}_D = \sum_{i=1}^{4} p_i^C\,|\psi_i^D\rangle\langle\psi_i^D|,
\label{eq:rhoBD}
\end{equation}
where $|\psi_i^B\rangle$ and $|\psi_i^D\rangle$ denote the eigenstates of Eqs.~(\ref{eq:psi1})--(\ref{eq:psi4}) evaluated at the hot $(B_{1h},B_{2h})$ and cold $(B_{1c},B_{2c})$ fields, respectively.

Over a complete cycle the internal energy returns to its initial value, so energy conservation requires
\[
\Delta U_{\text{cycle}} = W_{AB} + Q_h + W_{CD} + Q_c = 0.
\]
The net work output and the engine efficiency are defined as \cite{Peng2019, AbdRabbou2023, Khlifi2020}
\begin{align}
W_{\text{net}} &= -(W_{AB} + W_{CD}) = Q_h + Q_c, \label{eq:net_work}\\
\eta &= \frac{W_{\text{net}}}{Q_h} = 1 + \frac{Q_c}{Q_h}, \qquad 
\text{(with } W_{\text{net}}>0,\; Q_h>0\text{)}. \label{eq:efficiency_engine}
\end{align}
Here $W_{\text{net}}>0$ means the system delivers net work, a convention that matches common usage. For a two-level working substance the ideal Otto efficiency would be $\eta_{\text{Otto}} = 1 - \epsl_c/\epsl_h$, while the ultimate Carnot limit is $\eta_{\text{Carnot}} = 1 - T_c/T_h$. The signs of the various quantities determine the operating mode \cite{Kuznetsova2023, Ahadpour2021}: $W_{\text{net}}>0,\; Q_h>0,\; Q_c<0$ corresponds to a heat engine; $W_{\text{net}}<0,\; Q_h<0,\; Q_c>0$ corresponds to a refrigerator.

All thermodynamic quantities can be evaluated explicitly from the analytical expressions for $E_i$ and $p_i$. For example, the heat absorbed from the hot reservoir takes the closed form
\begin{equation}
Q_h = \sum_{i=1}^4 E_i^B \left( \frac{e^{-\beta_h E_i^B}}{Z_h} - \frac{e^{-\beta_c E_i^A}}{Z_c} \right),
\end{equation}
with $Z_{c,h}$ given by Eq.~\eqref{eq:partition} evaluated at the corresponding fields and temperatures. Although the sums involve four terms, the block-diagonal structure of the Hamiltonian allows them to be grouped into contributions from the two independent subspaces, which makes the numerical evaluation fast and transparent. No further simplification is needed for the purposes of this work.

\section{Performance analysis of the cycle}

\begin{figure}[htbp]
\centering
\setlength{\belowcaptionskip}{-10pt}
\includegraphics[width=0.9\linewidth]{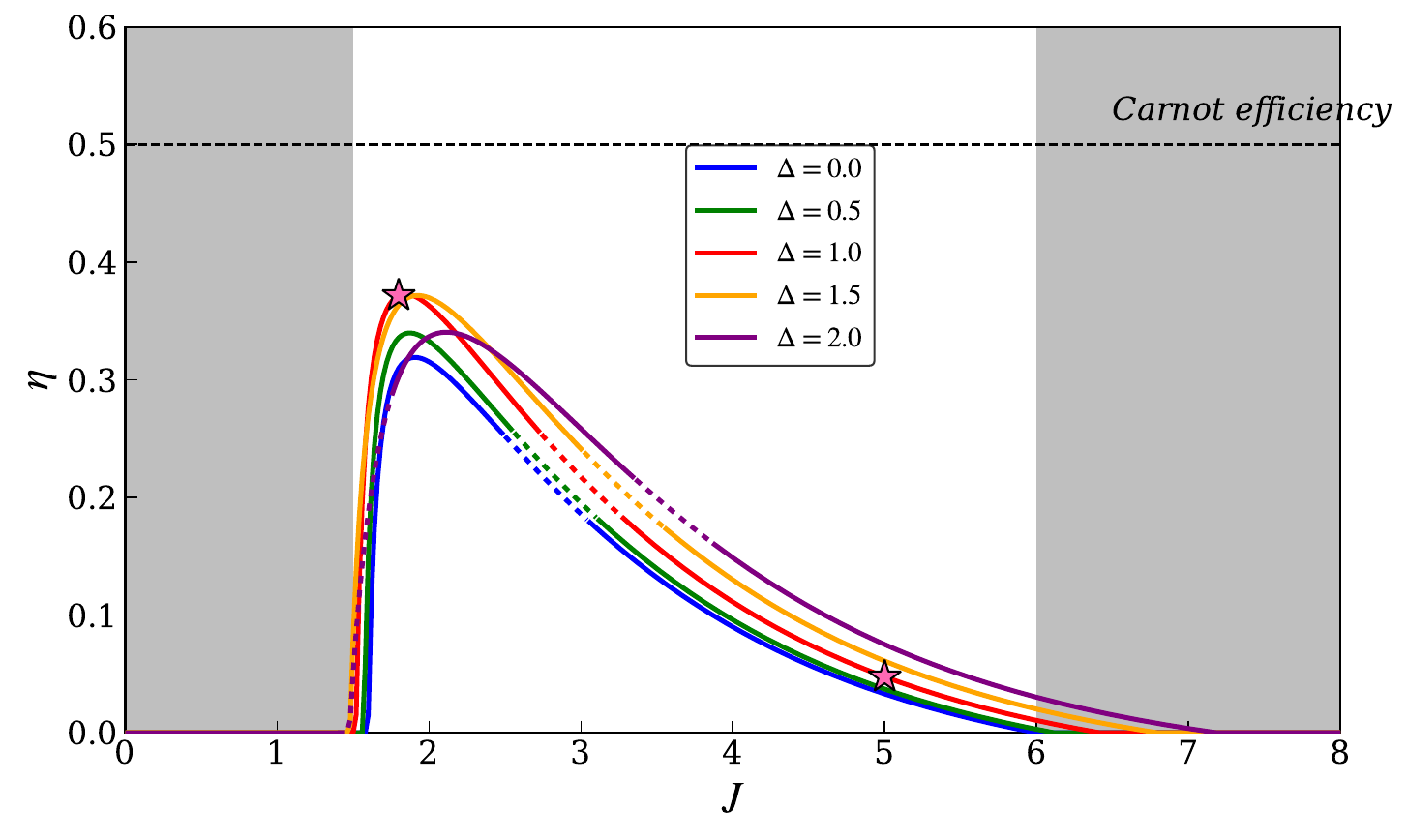}
\caption{Efficiency $\eta$ versus transverse coupling $J$ for several values of the anisotropy $\Delta$. Common parameters: $J_z=1.0$, $T_h=2.0$, $T_c=1.0$, $B_{1h}=3.0$, $B_{2h}=0.0$, $B_{1c}=0.8$, $B_{2c}=0.8$. The black dashed line marks the Carnot limit $\etac=0.5$. Shaded gray regions indicate parameter regimes where no net work can be extracted, i.e., where the heat engine does not operate in a valid engine mode. Dotted segments mark regions with level crossings, where normal heat-engine operation is also forbidden.}
\label{fig:eta_vs_J_Delta}
\end{figure}

\begin{figure}[htbp]
\centering
\includegraphics[width=1.0\linewidth]{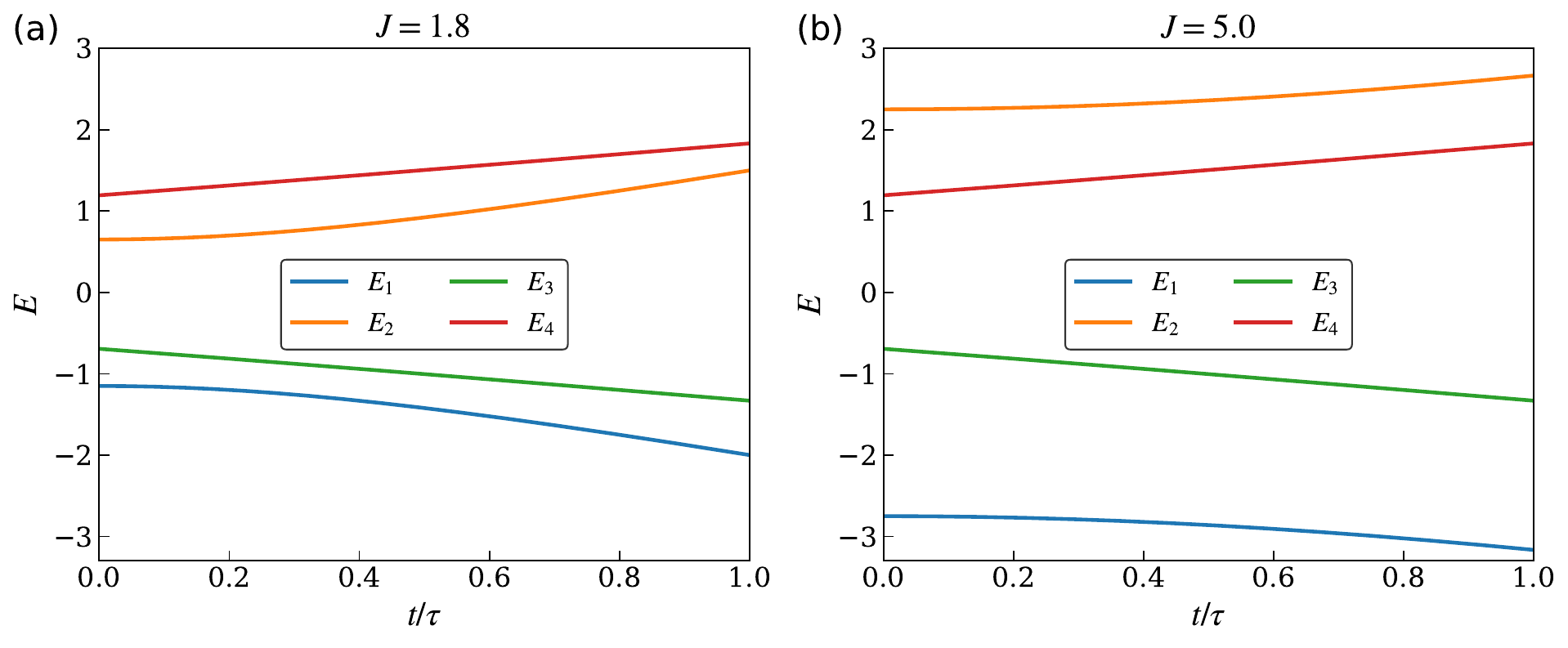}
\caption{Energy-level evolution during the linear time-dependent variation of the local magnetic fields, corresponding to the two star-marked points in Fig.~\ref{fig:eta_vs_J_Delta}. The solid lines from bottom to top at the initial time represent $E_1$, $E_3$, $E_2$, and $E_4$, respectively.}
\label{fig:energy_gap}
\end{figure}

\begin{figure}[htbp]
\centering
\includegraphics[width=\textwidth]{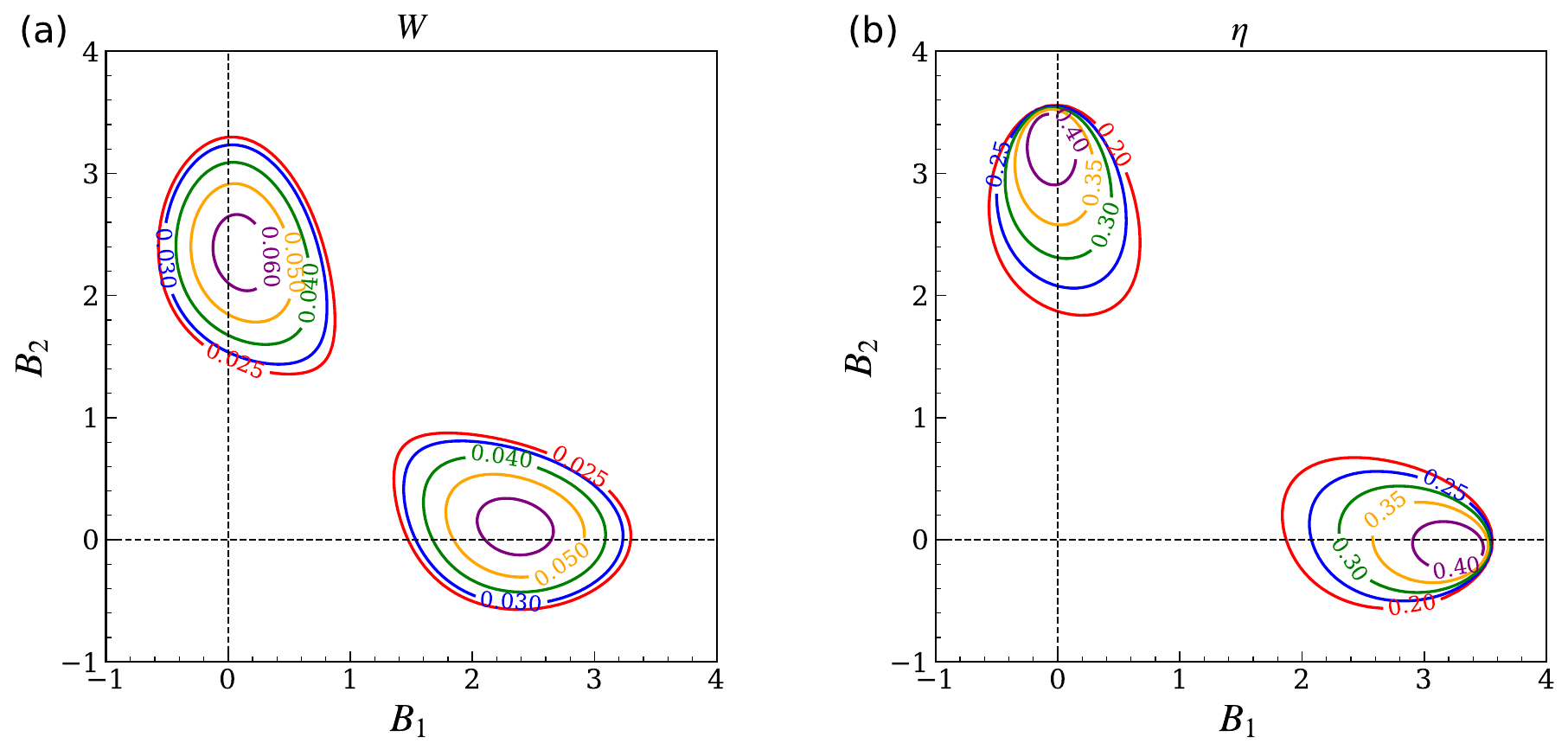}
\caption{Contour plots of net work $W_{\text{net}}$ and efficiency $\eta$ in the $(B_{1h},B_{2h})$ plane for $J_z=0.5$. Common parameters: $\Delta=1.0$, $J=2.0$, $T_h=2.0$, $T_c=1.0$, $(B_{1c},B_{2c})=(0.8,0.8)$. The black dashed lines indicate $B_{1h}=0$ or $B_{2h}=0$.}
\label{fig:W_eta_B1B2_Jz0.5}
\end{figure}

\begin{figure}[htbp]
\centering
\includegraphics[width=\textwidth]{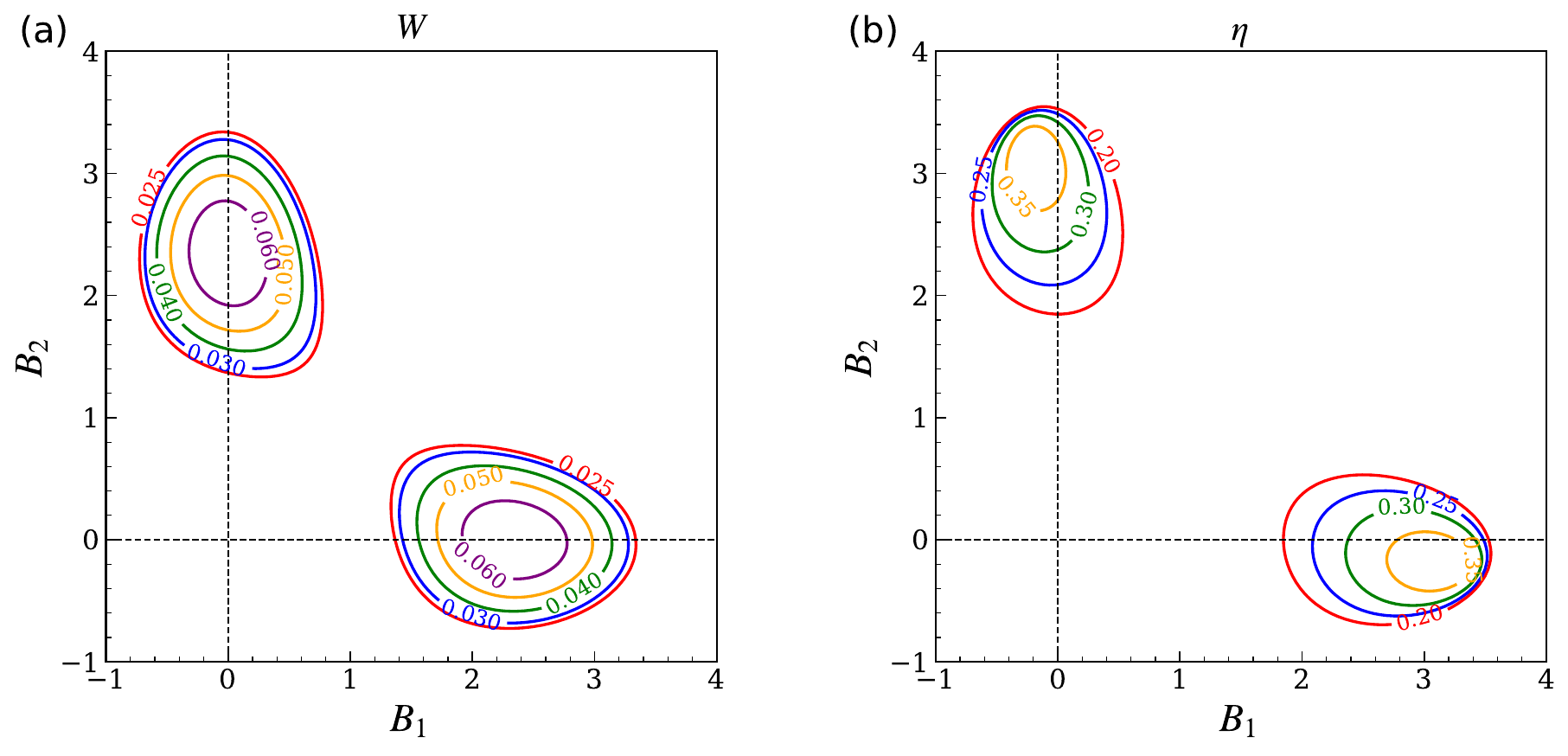}
\caption{Same as Fig.~\ref{fig:W_eta_B1B2_Jz0.5} but for $J_z=1.0$.}
\label{fig:W_eta_B1B2_Jz1.0}
\end{figure}

\begin{figure}[htbp]
\centering
\includegraphics[width=\linewidth]{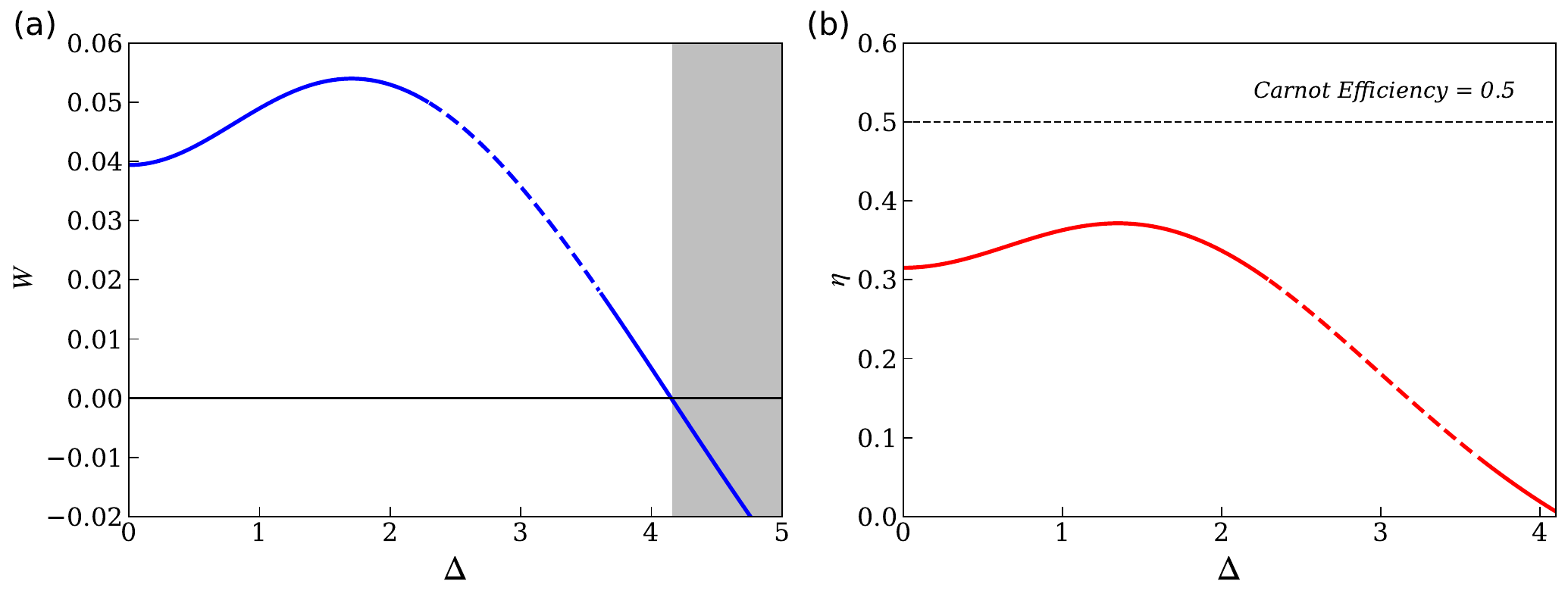}
\caption{Net work $W_{\text{net}}$ and efficiency $\eta$ as functions of the anisotropy $\Delta$. Parameters: $J_z=1.0$, $J=2.0$, $T_h=2.0$, $T_c=1.0$, $B_{1h}=3.0$, $B_{2h}=0.0$, $B_{1c}=0.8$, $B_{2c}=0.8$.}
\label{fig:W_eta_vs_Delta}
\end{figure}

With the exact eigenstates at hand, we now examine the performance of the quantum Otto engine. All simulations share the following baseline parameters: Boltzmann constant $k_B=1$, transverse coupling $J=2.0$, temperatures $T_h=2.0$ and $T_c=1.0$ (so that the Carnot efficiency is $\etac=0.5$). The cold-end magnetic fields are fixed at $(B_{1c},B_{2c})=(0.8,0.8)$, a choice that keeps the cold-end energy levels non-degenerate and ensures a well-defined thermal state as a stable reference for the cycle. The hot-end fields are varied over the ranges indicated in each figure. Figures~\ref{fig:eta_vs_J_Delta}--\ref{fig:local_work} illustrate how different parameters control the engine's behavior.

Figure~\ref{fig:eta_vs_J_Delta} shows the efficiency $\eta$ as a function of $J$ for several values of $\Delta$. The cold-end fields are $(B_{1c},B_{2c})=(0.8,0.8)$ and the hot-end fields are $(B_{1h},B_{2h})=(3.0,0.0)$. For a fixed $\Delta$, the thermal efficiency passes through four distinct regimes as $J$ is varied. For $0 \leq J \lesssim 1.5$, no net work can be extracted. In the range $1.5 \lesssim J \lesssim 2.0$, $\eta$ rises sharply to its peak value, after which it decreases gradually for $2.0 \lesssim J \lesssim 6.0$, and finally drops out of the valid operating regime for $J>6.0$. As the system moves from the isotropic $XY$ limit ($\Delta=0$) toward the anisotropic regime, the peak efficiency first increases and then decreases, attaining a maximum of about $0.38$ at $\Delta=1.0$. This peak efficiency is comparable to the values reported in recent superconducting quantum heat engine experiments \cite{Uusnakki2025a, Uusnakki2026}, which typically lie between $0.1$ and $0.4$ for similar temperature ratios.

These qualitative features are robust against the choice of the temperature ratio. With $T_c=1.0$ held fixed and $T_h/T_c$ varied over $1.5$--$3.0$, the peak efficiency stays in a narrow band at about $0.72$--$0.75$ of $\etac$, while lowering the temperature ratio shrinks the operating window in $J$ and makes a finite anisotropy $\Delta\gtrsim1$ indispensable for the engine mode; the details are given in Appendix~\ref{app:robustness}.

To characterize the level crossings analytically, we note from Eqs.~(\ref{eq:eig1})--(\ref{eq:eig2}) that degeneracies occur when two eigenvalues coincide. For the scans in Fig.~\ref{fig:eta_vs_J_Delta}, the crossings encountered along the driving path are those between $E_2$ and $E_4$: as $J$ increases from $1.8$ to $5.0$ for $\Delta=1.0$ (the two star-marked points), the energy-level evolution in Fig.~\ref{fig:energy_gap} shows that $E_2$ and $E_4$ meet inside the dotted region. Setting $E_2=E_4$ gives the crossing condition
\begin{equation}
\paramv - \paramu = J_z,
\label{eq:level_crossing}
\end{equation}
where $\paramu$ and $\paramv$ are evaluated with the instantaneous fields of the stroke in which the crossing occurs. Whenever this condition is approximately met along the path, the relevant energy gap closes, the adiabatic approximation breaks down, and normal engine operation is forbidden.

The anomalous efficiency drop for the isotropic case ($\Delta=0$) near $J \approx 1.8$ deserves special attention. In contrast to the anisotropic cases, the full $XY$ symmetry at $\Delta=0$ makes the energy spectrum highly susceptible to level crossings under the strongly asymmetric hot-end fields $(B_{1h}, B_{2h}) = (3.0, 0.0)$. Without the anisotropy gap, the adiabatic driving path can intersect degeneracy points in parameter space much earlier than it does for $\Delta > 0$. This premature level crossing locally invalidates the quantum adiabatic theorem and induces non-adiabatic transitions that severely degrade the work extraction capability, pushing the system out of the valid engine mode.

Level crossings occur during the field driving at $J\approx3.0$ and, for $\Delta=2.0$, at $J\approx1.5$; both invalidate the adiabatic approximation and break the normal engine operation. These parameter regions are indicated by dotted segments in the corresponding curves. For $1.0 \lesssim \Delta \lesssim 1.5$ and $1.8 \lesssim J \lesssim 2.8$, the efficiency exceeds $0.30$ and no level crossing occurs, demonstrating a robust operational window of coupling parameters for this quantum Otto heat engine. All curves lie strictly below the Carnot limit $\eta_C=0.5$, as required by the second law.

The non-monotonic behavior with $J$ originates from the competition between the energy-level compression factor and the thermal occupation of higher levels: very small $J$ leads to nearly degenerate eigenstates and poor work extraction, whereas excessively large $J$ reduces the relative change of the energy gaps during the adiabatic strokes, thereby suppressing $W_{\rm net}$ and $\eta$. To corroborate this interpretation, we explicitly evaluate the evolution of the individual energy levels during the adiabatic compression stroke. We assume that the local fields vary linearly in time as $B_{i}(t)=B_{ic}+\frac{t}{\tau}(B_{ih}-B_{ic})$ ($i=1,2$). For the two star-marked points in Fig.~\ref{fig:eta_vs_J_Delta}---the efficiency peak at $J=1.8$ and the low-efficiency large-$J$ case at $J=5.0$, both for $\Delta=1.0$---we compute the energy-level trajectories from the initial to the final state, as shown in Fig.~\ref{fig:energy_gap}. In the large-$J$ case the pairs $(E_1,E_3)$ and $(E_2,E_4)$ exhibit a nearly vanishing relative gap variation (the energy differences $E_3 - E_1$ and $E_4 - E_2$ remain essentially constant during the stroke), and the larger absolute energy separations make the relative gap variation even less pronounced.

Figures~\ref{fig:W_eta_B1B2_Jz0.5} and~\ref{fig:W_eta_B1B2_Jz1.0} display contour maps of $W_{\rm net}$ and $\eta$ in the $(B_{1h},B_{2h})$ plane for $J_z=0.5$ and $J_z=1.0$, respectively. In both panels the cold-end fields are fixed at $(B_{1c},B_{2c})=(0.8,0.8)$, and the couplings are set to $\Delta=1.0$, $J=2.0$. The contour lines are symmetric about the diagonal $B_{1h}=B_{2h}$, a direct consequence of the exchange symmetry of the Hamiltonian in Eq.~\eqref{eq:hamiltonian}. Moreover, level crossings are absent over most of the parameter space enclosed by the contours, making these regions valid for operating the Otto heat engine.

For $J_z=0.5$, the maximum net work reaches $W_{\rm net}\approx0.06$ in a broad region centered near $(B_{1h},B_{2h})\approx(2.4,0.2)$, while the efficiency reaches its peak $\eta\approx0.45$ around $(3.3,-0.1)$. The region where $\eta>0.35$ covers a substantial fraction of the parameter space, indicating that a weak longitudinal coupling $J_z$ favors both high work output and high efficiency.

When $J_z$ is increased to $1.0$ (Fig.~\ref{fig:W_eta_B1B2_Jz1.0}), the accessible work does not change appreciably, but the peak efficiency is suppressed to $\eta<0.40$. Physically, a larger $J_z$ introduces an extra energy cost for spin alignment along the $z$-axis, which reduces the effective compression ratio of the working substance and thereby lowers the extractable work. Operating the engine at small $J_z$ is therefore preferable for approaching the Carnot efficiency.

Figure~\ref{fig:W_eta_vs_Delta} plots $W_{\rm net}$ and $\eta$ against $\Delta$ for fixed fields $(B_{1c},B_{2c})=(0.8,0.8)$ and $(B_{1h},B_{2h})=(3.0,0.0)$, with $J=2.0$, $J_z=1.0$. Both quantities display a clear non-monotonic dependence on $\Delta$. For $\Delta\lesssim0.5$, the $XY$ coupling is nearly isotropic and the energy-level modulation during the cycle is insufficient; $W_{\rm net}$ stays around $0.04$ and $\eta$ around $0.32$. As $\Delta$ increases, $W_{\rm net}$ rises noticeably, reaching a maximum of $\approx0.055$ at $\Delta\approx2.0$, while $\eta$ peaks at $\approx0.38$ near $\Delta\approx1.5$. Beyond these optimal values, a further increase of $\Delta$ leads to a gradual decay: at $\Delta=4.2$ the work output vanishes, signaling that the system has left the valid operating regime of the Otto heat engine.

This behavior can be understood from the eigenenergies in Eqs.~\eqref{eq:eig1}--\eqref{eq:eig2}. When $\Delta$ is large, the gaps $E_3-E_1$ and $E_4-E_2$ are dominated by $\Delta$, which suppresses their relative change under the variation of the magnetic fields and thus reduces the net work. An optimal $\Delta$ therefore balances the generation of sufficient anisotropy to lift degeneracies against an excessive stiffening of the energy spectrum. Level crossings occur for $2.3\lesssim\Delta\lesssim3.6$, making this interval unsuitable as a valid parameter window for the heat engine; fortunately, this range lies away from the region where the net work and thermal efficiency attain their maxima.

To place our findings in context, it is helpful to compare the maximum efficiency $\eta \approx 0.45$ (reaching $90\%$ of the Carnot limit) with recent results for related spin-based quantum heat engines. Kuznetsova \textit{et al.} \cite{Kuznetsova2023}, for instance, studied an Otto engine with an XYZ working medium that includes DM and KSEA interactions, and reported optimal efficiencies typically below $0.40$ for comparable temperature ratios. Our result shows that independent local magnetic fields can yield a superior compression ratio and higher efficiency without requiring complex spin-orbit couplings. Furthermore, while recent studies of Stirling engines (e.g., Xiao \textit{et al.} \cite{Xiao2026} and Tang \textit{et al.} \cite{Tang2026}) have found that the Heisenberg exchange $J$ acts as the primary control parameter for mode transitions (among engine, heater, and refrigerator), our Otto cycle analysis reveals a different mechanism: the anisotropy $\Delta$ and the asymmetry of the local fields together govern the valid operational window, avoiding the level-crossing-induced breakdowns that afflict Stirling cycles in the strong-coupling regime. This highlights the unique advantage of local field tuning in anisotropic Otto engines.

\section{Local work analysis and the role of quantum entanglement}
\label{sec:micro}

To gain insight into the individual roles of the two qubits and their interaction during the cycle, we introduce the concept of local work \cite{Peng2019, Thomas2011}, which has been widely used to analyze energy partitioning in composite quantum systems and to interpret the work statistics of quantum measurement engines \cite{Bresque2025}. We decompose the Hamiltonian into three parts: the magnetic-field terms $H_1$ and $H_2$ acting on qubits 1 and 2, and the interaction term $H_{\rm int}$. The single-qubit terms change with the magnetic-field strengths at different stages; for the adiabatic compression and expansion steps they read
\begin{align}
\hat{H}_{1c} &= B_{1c} \hat{S}_1^z,\quad \hat{H}_{1h} = B_{1h} \hat{S}_1^z,\\
\hat{H}_{2c} &= B_{2c} \hat{S}_2^z,\quad \hat{H}_{2h} = B_{2h} \hat{S}_2^z.
\end{align}

We then analyze separately the work that each of these three terms contributes throughout the full Otto cycle. As in the total work calculation above, work from an individual term is produced only during the adiabatic compression step $AB$ and the adiabatic expansion step $CD$.

We first obtain the density matrix for each state: the equilibrium states $\hat{\rho}_A$ and $\hat{\rho}_C$ are the Gibbs states computed from the formulas of Sec.~\ref{sec: theory and model}, while the nonequilibrium states $\hat{\rho}_B$ and $\hat{\rho}_D$ are constructed from Eq.~(\ref{eq:rhoBD}). The work originating from $H_1$ during step $AB$, for example, is
\begin{equation}
    \begin{split}
        w_{1,AB} &= \operatorname{Tr}\left[\hat{\rho}_{A}(\hat{H}_{1c}\otimes \hat{I}_2)-\hat{\rho}_{B}(\hat{H}_{1h}\otimes \hat{I}_2)\right] \\
        &= \operatorname{Tr}(\hat{\rho}_{A1}\hat{H}_{1c}-\hat{\rho}_{B1}\hat{H}_{1h}),
    \end{split}
\end{equation}
where $\hat{\rho}_{A1}=\operatorname{Tr}_2(\hat{\rho}_{A})$ is the reduced density matrix of $\hat{\rho}_A$ for qubit 1, and similarly for state $B$. We adopt this trace-based partition because it directly quantifies the work done by each physically distinct part of the Hamiltonian without introducing additional symmetrization prescriptions. This approach is consistent with the standard operational definition of local work in composite quantum systems, where each subsystem's contribution is evaluated via its reduced dynamics \cite{Thomas2011, Peng2019}. The three contributions to the work output then read
\begin{align}
    w_1 = w_{1,AB}+w_{1,CD} &= \operatorname{Tr}\left[(\hat{\rho}_{A1}-\hat{\rho}_{D1})\hat{H}_{1c}+(\hat{\rho}_{C1}-\hat{\rho}_{B1})\hat{H}_{1h}\right], \\
    w_2 = w_{2,AB}+w_{2,CD} &= \operatorname{Tr}\left[(\hat{\rho}_{A2}-\hat{\rho}_{D2})\hat{H}_{2c}+(\hat{\rho}_{C2}-\hat{\rho}_{B2})\hat{H}_{2h}\right], \\
    w_{\rm int} = w_{{\rm int},AB}+w_{{\rm int},CD} &= \operatorname{Tr}\left[(\hat{\rho}_{A}+\hat{\rho}_{C}-\hat{\rho}_{B}-\hat{\rho}_{D})\hat{H}_{\rm int}\right].
\end{align}
The total net work is $W_{\rm{net}} = w_1 + w_2 + w_{\rm int}$. The sign of each $w$ directly reveals its thermodynamic function: $w > 0$ means that the corresponding term acts as a heat engine (delivering work), while $w < 0$ indicates that it operates as a refrigerator (absorbing work).

\begin{figure}[htbp]
\centering
\includegraphics[width=0.9\linewidth]{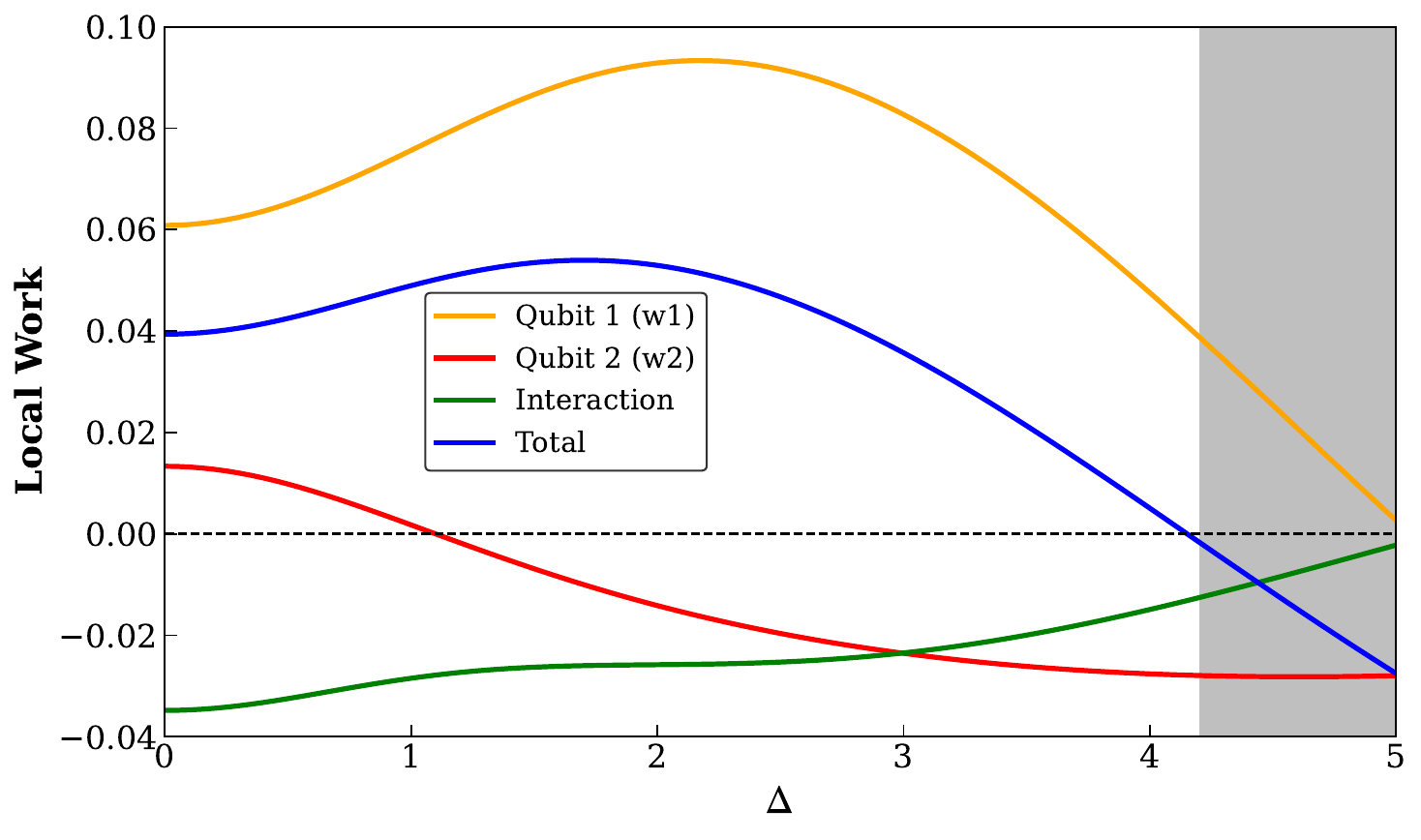}
\caption{Local work $w_1$ and $w_2$ of the two qubits, together with the interaction work $w_{\rm int}$, as functions of $\Delta$. Parameters are the same as in Fig.~\ref{fig:W_eta_vs_Delta}. The blue line shows the total net work; the black dashed line marks zero work.}
\label{fig:local_work}
\end{figure}

Figure~\ref{fig:local_work} displays $w_1$, $w_2$, and $w_{\rm int}$ as functions of $\Delta$ for the same parameters as in Fig.~\ref{fig:W_eta_vs_Delta}. A clear functional differentiation emerges: $w_1$ is strictly positive throughout the valid range, $w_2$ is positive for $\Delta\lesssim 1$ but becomes negative at larger $\Delta$, and $w_{\rm int}$ remains negative everywhere.

In detail, $w_1 = +0.06$ at $\Delta=0$ and increases continuously as $\Delta$ grows from $0$ to $2$. For $2\lesssim\Delta\lesssim 3$, $w_1$ stays at a relatively high level around $+0.09$ and then decreases. By contrast, $w_2$ decreases monotonically from $+0.013$ to zero and further toward $-0.03$, while $w_\mathrm{int}$ rises monotonically from $-0.035$ without ever reaching zero in the valid working regime. Thus qubit 1 consistently acts as a heat engine, delivering positive work, whereas qubit 2 first acts as a heat engine and then becomes a refrigerator. The interaction term absorbs work throughout the cycle.

To understand this functional differentiation quantitatively without resorting to numerical diagonalization at every step, we examine the analytical structure of the eigenstates and the reduced density matrices. The critical transition of qubit 2 from a heat engine ($w_2 > 0$) to a refrigerator ($w_2 < 0$) near $\Delta \approx 1$ is rooted in the level mixing within the $\{|01\rangle, |10\rangle\}$ subspace. According to Eq.~(\ref{eq:hamiltonian}), the effective Hamiltonian in this subspace is governed by the competition between the transverse coupling $J$ and the field difference $\etab$. As $\Delta$ increases, the eigenstates $|\psi_3\rangle$ and $|\psi_4\rangle$ in the $\{|00\rangle, |11\rangle\}$ subspace become increasingly dominated by the anisotropy term, effectively decoupling from the thermal population dynamics of the single-spin excitations. At the critical point where $\Delta \sim J$, the energy gap between the ground state and the first excited state undergoes an avoided crossing or significant restructuring, modifying the thermal occupation probabilities $p_i$.

Quantitatively, the reduced density matrix of qubit 2, $\hat{\rho}_2 = \operatorname{Tr}_1(\hat{\rho})$, has diagonal elements that depend strictly on the populations of $|01\rangle$ and $|11\rangle$. Once $\Delta$ exceeds the threshold, the thermal weight shifts heavily toward the $|00\rangle$ and $|11\rangle$ states because of the large energy penalty $\paramu \approx \Delta$, suppressing the population of $|01\rangle$. The expectation value of $\hat{H}_{2h} = B_{2h}\hat{S}_2^z$ then becomes less sensitive to the adiabatic variation, and the work integral $w_2 = \operatorname{Tr}[(\hat{\rho}_{C2}-\hat{\rho}_{B2})\hat{H}_{2h}]$ changes sign. This analytical mechanism confirms that $H_{\rm int}$ acts as an internal work reservoir, absorbing work to pump qubit 2 against its local field gradient. The behavior is reminiscent of the spin-localization effects induced by inhomogeneous fields in non-extensive thermodynamic frameworks \cite{Han2026} and of the dominance of local system-reservoir couplings in determining subsystem thermodynamic roles \cite{Mashhor2025}.

The strongly asymmetric local magnetic fields in the hot stroke ($B_{1h}=3.0$, $B_{2h}=0.0$) are the direct cause of this functional differentiation. We stress that this asymmetry is introduced as an external control protocol; it does not mimic any intrinsic material asymmetry. Although such an extreme field imbalance may entail additional control overhead in experiments, it provides a transparent theoretical limit in which the fundamental mechanisms of work partitioning can be exposed. In practice, minor crosstalk between local fields is unavoidable, but a small perturbation of the form $B_{2h} = \epsilon B_{1h}$ with $\epsilon \ll 1$ would merely shift the critical value of $\Delta$ at which $w_2$ changes sign, leaving the overall asymmetric redistribution mechanism qualitatively intact. Qubit 1 experiences a larger field variation ($B_{1h}>B_{1c}$), which in a naive single-spin picture would make it a work absorber. However, the spin--spin couplings $J$ and $\Delta$ induce level mixing that effectively inverts the role of qubit 1, allowing it to extract net work. Conversely, the lower field on qubit 2 ($B_{2h}<B_{2c}$) would normally favor engine operation, but the coupling to qubit 1 reduces its output and even forces it into a refrigerator role.

The interaction term does more than modify the energy-level structure: it also absorbs part of the external work input during the cycle. This absorbed work does not diminish the system's ability to deliver work to the environment; it simply redistributes the output. In the present case, the redistribution enhances $w_1$, the work output of qubit 1. This suggests that tailored thermal contacts between one qubit and other systems, together with the anisotropic XYZ working substance, could lead to improved performance. More broadly, the idea that interaction terms can be deliberately engineered to act as internal energy buffers may find applications beyond single quantum heat engines---for instance, in the design of multi-qubit quantum batteries or many-body thermal machines where energy redistribution among subsystems is desirable \cite{Campaioli2017, Ferraro2018}.

\subsection{Quantum entanglement along the cycle}

From the perspective of quantum information, bipartite entanglement between the two qubits offers additional insight into the microscopic operation of the heat engine. For a two-qubit system the concurrence $\mathcal{C}$ is a faithful measure of entanglement \cite{Wootters1998}. The Gibbs states $A$ and $C$, given by Eqs.~(\ref{eq:rho11})--(\ref{eq:rho23}), have an X-state form (block-diagonal with non-zero elements only on the diagonal and anti-diagonal). The nonequilibrium states $B$ and $D$ of Eq.~(\ref{eq:rhoBD}) share this structure, because each eigenstate of Eqs.~(\ref{eq:psi1})--(\ref{eq:psi4}) lives entirely within one of the two subspaces $\{|00\rangle,|11\rangle\}$ and $\{|01\rangle,|10\rangle\}$; all four density matrices are therefore X-states. For such X-states the general Wootters formula reduces to the compact analytical expression \cite{Wootters1998}
\begin{equation}
\mathcal{C} = 2 \max\Bigl(0,\; |\rho_{14}| - \sqrt{\rho_{22}\rho_{33}},\; |\rho_{23}| - \sqrt{\rho_{11}\rho_{44}}\Bigr).
\label{eq:concurrence_X}
\end{equation}
This simplification is a significant advantage: it allows the concurrence to be evaluated directly from the closed-form density matrix elements without further numerical diagonalization. Because $\rho_{14} \propto -\Delta$ and $\rho_{23} \propto -J$ are negative in the thermal equilibrium state (see Eqs.~(\ref{eq:rho14}) and (\ref{eq:rho23})), their absolute values directly quantify the competing strengths of the anisotropy and the exchange coupling in generating quantum correlations. The two terms $|\rho_{14}| - \sqrt{\rho_{22}\rho_{33}}$ and $|\rho_{23}| - \sqrt{\rho_{11}\rho_{44}}$ reflect entanglement contributions from the two independent subspaces $\{|00\rangle,|11\rangle\}$ and $\{|01\rangle,|10\rangle\}$, respectively.

\begin{figure}[htbp]
\centering
\includegraphics[width=0.9\linewidth]{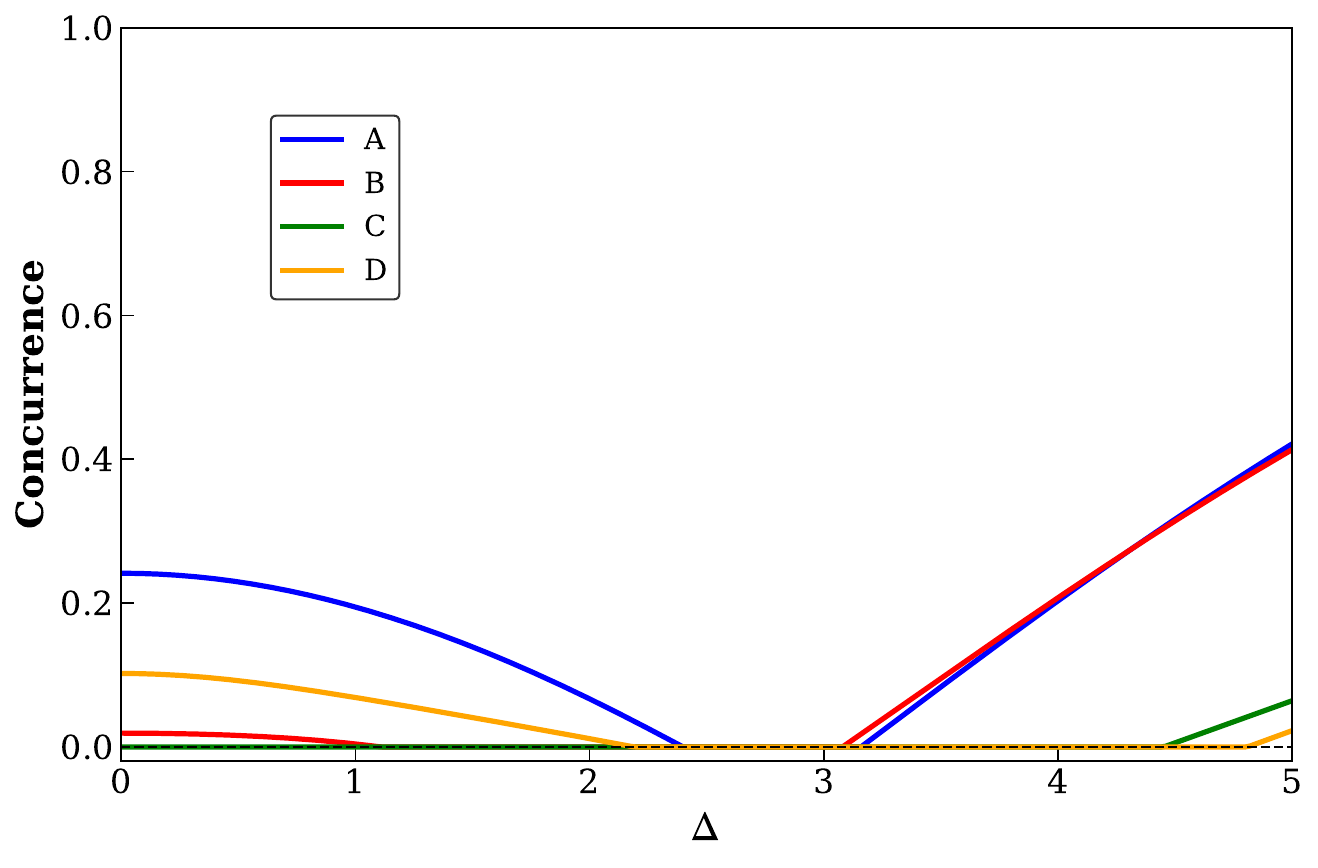}
\caption{Concurrence $\mathcal{C}$ between the two qubits as a function of $\Delta$ at the four stages $A$, $B$, $C$, and $D$ of the Otto cycle. Parameters are the same as in Fig.~\ref{fig:W_eta_vs_Delta}. The black dashed line indicates the boundary of zero entanglement.}
\label{fig:concurrence}
\end{figure}

Figure~\ref{fig:concurrence} shows the concurrence of the four characteristic states, evaluated with Eq.~(\ref{eq:concurrence_X}). The two competing terms of this expression generate two distinct entanglement sectors. Since $J > B_{1c}+B_{2c} = 1.6$, the cold-end states $A$ and $D$ are entangled already at $\Delta=0$ through the $|\rho_{23}|$ contribution; this contribution is suppressed with increasing $\Delta$ and vanishes at $\Delta\simeq2.2$--$2.4$. The $|\rho_{14}|$ contribution takes over at $\Delta\simeq3.1$, where the concurrence of $A$ and $B$ re-emerges and grows rapidly, while the hot-end state $C$ remains disentangled over most of the range ($J < B_{1h}+B_{2h} = 3.0$; it is born only at $\Delta\simeq4.5$).

Notably, in the intermediate window $\Delta\simeq2.2$--$3.1$ the engine operates with all four states disentangled, while the re-emergence of the concurrence beyond $\Delta\simeq3.1$ coincides with the level-crossing window in which the performance degrades. The concurrence therefore serves as a witness of the spectral restructuring driven by $\Delta$---marking the low-$\Delta$ operating regime through the cold-end entanglement asymmetry and signalling the crossing window through its re-emergence---rather than a resource for the work extraction. This perspective aligns with recent work that treats quantum correlations as indicators of favorable energy-level configurations for work extraction \cite{Bresque2025, Han2026}.

\section{Conclusion}

We have investigated a quantum Otto heat engine based on an anisotropic two-qubit Heisenberg XYZ model, with independent local magnetic fields applied to each spin. A systematic study of how the longitudinal coupling $J_z$, the anisotropy $\Delta$, the transverse coupling $J$, and the local fields regulate the net work output and efficiency has been presented.

The analysis shows that reducing the longitudinal coupling can significantly enhance both the maximum work output and the peak efficiency. The anisotropy parameter exhibits a non-monotonic effect: below a threshold the engine barely operates, the performance first improves to an optimal peak at moderate anisotropy, and it then decays slowly as the anisotropy increases further. An analytical condition for the occurrence of level crossings, $\paramu = \paramv$, has been derived, providing a clear operational boundary for the valid engine regime.

From the viewpoint of local work, a robust asymmetric differentiation of thermodynamic roles between the two qubits emerges: for the chosen field configuration, qubit 1 consistently acts as a heat engine ($w_1>0$), whereas qubit 2 functions either as a heat engine ($w_2>0$) or as a refrigerator ($w_2<0$), depending on $\Delta$. This difference arises from the interplay between the local field asymmetry and the anisotropic spin--spin couplings. Equally important, the interaction term absorbs part of the work during the cycle, which in turn enhances the work output of qubit 1, effectively acting as an internal energy redistribution mechanism. This ``energy redistributor'' concept may be transferable to other quantum devices, such as multi-qubit quantum batteries or many-body thermal machines, where interactions could be engineered to buffer and redistribute energy among subsystems \cite{Campaioli2017, Ferraro2018}.

By quantifying bipartite entanglement through the X-state concurrence formula, the entanglement dynamics have been related to the engine operation: the two threshold-like contributions $|\rho_{23}|\propto J$ and $|\rho_{14}|\propto\Delta$ mark the low-$\Delta$ operating regime and the level-crossing window, respectively. The engine operates over an intermediate window with vanishing entanglement at all four stages, while the re-emergent concurrence accompanies the degradation of the performance. The entanglement asymmetry between the cold and hot strokes is thus a \emph{witness} of the underlying energy-level restructuring rather than an independent thermodynamic resource or a cause of the efficiency enhancement.

In summary, the longitudinal coupling, the anisotropy, and the local magnetic fields all play critical roles in tuning the efficiency and performance of this quantum heat engine. On the microscopic level, the two qubits perform distinct thermodynamic functions, the interaction term redistributes energy throughout the cycle, and the entanglement between the strokes serves as a witness of the spectral restructuring that governs the engine operation rather than a resource for it.

Future work can extend this analysis to include non-adiabatic effects and quantum friction, to examine the role of quantum coherence during the cycle, or to consider finite-time operation, which would be more directly relevant to experiments. It would be particularly interesting to explore how the local magnetic field tuning strategy proposed here could be adapted to autonomous quantum heat engine architectures \cite{Rasola2025, Uusnakki2026}, and to quantify the entanglement asymmetry in measurement-driven or feedback-controlled engines \cite{Bresque2025}. Extending the analysis to incorporate non-extensive statistical effects \cite{Han2026} and squeezed reservoir engineering \cite{Zhu2026} would further deepen the understanding of quantum correlations in nonequilibrium thermodynamics.

\section*{Acknowledgments}

This work was supported by the Natural Science Foundation of Xinjiang Uygur Autonomous Region (Grant No.~2023D01A42 and 2023D01B50), the Doctoral (Postdoctoral) Research Startup Foundation of Xinjiang Normal University (Grant No.~XJNUZBS2433 and XJNUZBS2414), the National Natural Science Foundation of China (Grant No.~12405025), the ``Tianchi Talents'' Talent Introduction Program of the Xinjiang Uygur Autonomous Region, the Open Research Project of the Scientific Research Innovation Platform of Xinjiang Normal University (Grant No.~XJDX20222-2026-02), and the Teaching Research and Reform Project of Xinjiang Normal University (Grant No.~SDJG2026-37). We thank the Xinjiang Key Laboratory of Luminescent Minerals and Optical Functional Materials at the School of Physics and Electronic Engineering for technical support.

\appendix
\setcounter{figure}{0}
\renewcommand{\thefigure}{A\arabic{figure}}

\section{Exact eigenstates derivation and impact of normalization correction}
\label{app:eigen_derivation}

The Pauli matrices are defined as
\begin{equation}
\hat{\sigma}_x = \begin{pmatrix} 0 & 1 \\ 1 & 0 \end{pmatrix},\quad
\hat{\sigma}_y = \begin{pmatrix} 0 & -i \\ i & 0 \end{pmatrix},\quad
\hat{\sigma}_z = \begin{pmatrix} 1 & 0 \\ 0 & -1 \end{pmatrix}. \tag{A1}
\end{equation}

With the spin operators $\hat{S}_i^\alpha = \frac{1}{2}\hat{\sigma}_i^\alpha$, inserting them into Eq.~\eqref{eq:H_original} and evaluating the tensor products yields the Hamiltonian matrix
\begin{equation}
\hat{H} = \begin{pmatrix}
\frac{J_z}{4} + \frac{B_1+B_2}{2} & 0 & 0 & \frac{\Delta}{2} \\[4pt]
0 & -\frac{J_z}{4} + \frac{B_1-B_2}{2} & \frac{J}{2} & 0 \\[4pt]
0 & \frac{J}{2} & -\frac{J_z}{4} - \frac{B_1-B_2}{2} & 0 \\[4pt]
\frac{\Delta}{2} & 0 & 0 & \frac{J_z}{4} - \frac{B_1+B_2}{2}
\end{pmatrix}. \tag{A2}
\end{equation}

Solving $\det(H - \lambda I)=0$ leads to the two block-diagonal conditions
\begin{equation}
\det\begin{pmatrix}
\frac{J_z}{4} + \frac{\epsl}{2} - \lambda & \frac{\Delta}{2} \\
\frac{\Delta}{2} & \frac{J_z}{4} - \frac{\epsl}{2} - \lambda
\end{pmatrix} = 0, \qquad
\det\begin{pmatrix}
-\frac{J_z}{4} + \frac{\etab}{2} - \lambda & \frac{J}{2} \\
\frac{J}{2} & -\frac{J_z}{4} - \frac{\etab}{2} - \lambda
\end{pmatrix} = 0, \tag{A3}
\end{equation}
with $\epsl = B_1+B_2$ and $\etab = B_1-B_2$. Expanding these determinants gives quadratic equations whose solutions are
\begin{equation}
\lambda_{1,2} = -\frac{J_z}{4} \pm \frac{1}{2}\sqrt{\etab^2 + J^2},\qquad
\lambda_{3,4} = \frac{J_z}{4} \pm \frac{1}{2}\sqrt{\epsl^2 + \Delta^2}. \tag{A4}
\end{equation}
The corresponding normalized eigenstates are given in Eqs.~\eqref{eq:psi1}--\eqref{eq:psi4} of the main text.

To demonstrate the importance of correct normalization, we compare the internal energy computed with our normalized eigenstates to that obtained from the unnormalized states of Ref.~\cite{Abliz2006}. It is crucial to note that the canonical partition function $Z = \operatorname{Tr}(e^{-\beta \hat{H}}) = \sum_i e^{-\beta E_i}$ depends strictly on the energy eigenvalues---which are available in closed form in Eq.~(A4)---and is therefore completely independent of the eigenvector normalization. However, if unnormalized eigenvectors $|\phi_i\rangle$ are used to construct the thermal density matrix as $\hat{\rho}^{\,\prime} = \sum_i p_i |\phi_i\rangle\langle\phi_i|$ without the normalization denominator $\langle\phi_i|\phi_i\rangle$, the trace of the resulting matrix deviates from unity, $\operatorname{Tr}(\hat{\rho}^{\,\prime}) = \sum_i p_i \langle\phi_i|\phi_i\rangle \neq 1$, and all thermodynamic expectation values computed from $\hat{\rho}^{\,\prime}$, including the internal energy, acquire a systematic bias. The spectral projectors must therefore be written as $|\phi_i\rangle\langle\phi_i| / \langle\phi_i|\phi_i\rangle$, or equivalently the states must be normalized as in Eqs.~\eqref{eq:psi1}--\eqref{eq:psi4} of the main text. We emphasize that this is a property of the density-matrix construction, not of the partition function itself; the correct normalization is essential for quantitative accuracy in evaluating thermodynamic performance.

\section{Derivation of the closed-form thermal density matrix}
\label{app:rho_derivation}

The block-diagonal structure of the Hamiltonian allows each $2\times 2$ sub-block to be exponentiated independently. Consider the subspace spanned by $\{|00\rangle,|11\rangle\}$. The Hamiltonian in this subspace is
\[
\hat{H}_A = \frac{J_z}{4} \hat{I}_2 + \frac{1}{2}\begin{pmatrix} \epsl & \Delta \\ \Delta & -\epsl \end{pmatrix}.
\]
Writing $\hat{H}_A = \frac{J_z}{4} \hat{I}_2 + \vec{v}\cdot\vec{\sigma}$ with $\vec{v} = \frac{1}{2}(\Delta, 0, \epsl)$ and $|\vec{v}| = \paramu/2$, the exponential $e^{-\beta \hat{H}_A}$ can be evaluated using the identity $e^{\alpha \hat{I} + \vec{v}\cdot\vec{\sigma}} = e^{\alpha}[\cosh(|\vec{v}|)\hat{I} + \frac{\vec{v}\cdot\vec{\sigma}}{|\vec{v}|}\sinh(|\vec{v}|)]$. Dividing by $Z$ gives the matrix elements in the $\{|00\rangle,|11\rangle\}$ subspace:
\begin{align*}
\rho_{11} &= \frac{e^{-\beta J_z/4}}{Z} \left[ \cosh\left(\frac{\beta \paramu}{2}\right) - \frac{\epsl}{\paramu} \sinh\left(\frac{\beta \paramu}{2}\right) \right], \\
\rho_{44} &= \frac{e^{-\beta J_z/4}}{Z} \left[ \cosh\left(\frac{\beta \paramu}{2}\right) + \frac{\epsl}{\paramu} \sinh\left(\frac{\beta \paramu}{2}\right) \right], \\
\rho_{14} = \rho_{41} &= -\frac{e^{-\beta J_z/4}}{Z} \cdot \frac{\Delta}{\paramu} \sinh\left(\frac{\beta \paramu}{2}\right).
\end{align*}
Similarly, for the subspace $\{|01\rangle,|10\rangle\}$ the sub-Hamiltonian is
\[
\hat{H}_B = -\frac{J_z}{4} \hat{I}_2 + \frac{1}{2}\begin{pmatrix} \etab & J \\ J & -\etab \end{pmatrix},
\]
with $|\vec{v}| = \paramv/2$. The corresponding matrix elements are
\begin{align*}
\rho_{22} &= \frac{e^{\beta J_z/4}}{Z} \left[ \cosh\left(\frac{\beta \paramv}{2}\right) - \frac{\etab}{\paramv} \sinh\left(\frac{\beta \paramv}{2}\right) \right], \\
\rho_{33} &= \frac{e^{\beta J_z/4}}{Z} \left[ \cosh\left(\frac{\beta \paramv}{2}\right) + \frac{\etab}{\paramv} \sinh\left(\frac{\beta \paramv}{2}\right) \right], \\
\rho_{23} = \rho_{32} &= -\frac{e^{\beta J_z/4}}{Z} \cdot \frac{J}{\paramv} \sinh\left(\frac{\beta \paramv}{2}\right).
\end{align*}
The partition function $Z = \operatorname{Tr}(e^{-\beta \hat{H}})$ is obtained by summing the traces of $e^{-\beta \hat{H}_A}$ and $e^{-\beta \hat{H}_B}$, which recovers Eq.~\eqref{eq:partition}. This closed-form density matrix is fully equivalent to the eigenstate expansion and provides a convenient route for the numerical evaluation of local observables.

\section{Robustness against the temperature ratio}
\label{app:robustness}

\begin{figure}[htbp]
\centering
\includegraphics[width=\textwidth]{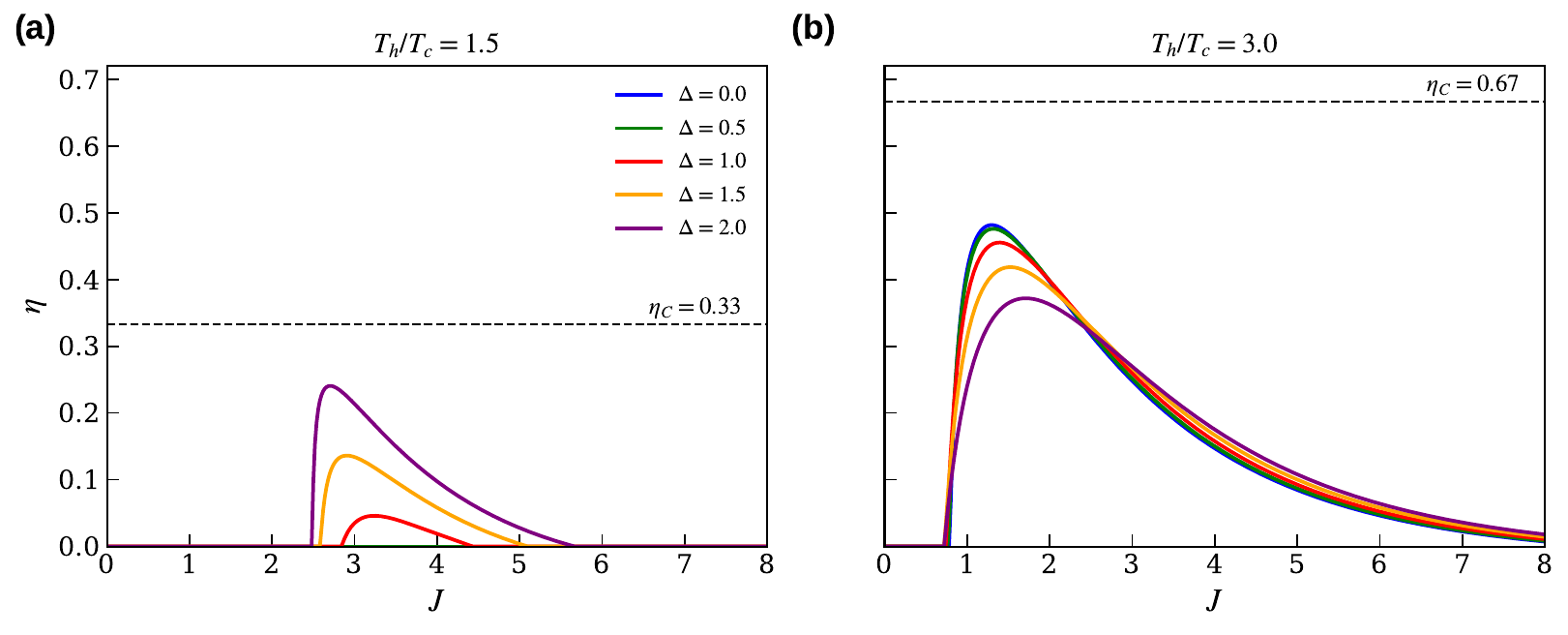}
\caption{Efficiency $\eta$ versus $J$ for $\Delta=0,0.5,1.0,1.5,2.0$ at (a) $T_h/T_c=1.5$ and (b) $T_h/T_c=3.0$, with $T_c=1.0$ fixed. The remaining parameters are those of the main text: $J_z=1.0$, $(B_{1c},B_{2c})=(0.8,0.8)$, $(B_{1h},B_{2h})=(3.0,0.0)$. The black dashed lines mark the Carnot efficiency $\etac=1-T_c/T_h$.}
\label{fig:eta_ratio}
\end{figure}

In the main text the temperatures are fixed to $T_h=2.0$ and $T_c=1.0$. Here we assess the robustness of the performance against the temperature ratio, keeping $T_c=1.0$ and the field configuration $(B_{1c},B_{2c})=(0.8,0.8)$, $(B_{1h},B_{2h})=(3.0,0.0)$ unchanged, and varying $T_h/T_c$ from $1.5$ to $3.0$. Figure~\ref{fig:eta_ratio} shows $\eta$ as a function of $J$ for $\Delta=0,0.5,1.0,1.5,2.0$ at $T_h/T_c=1.5$ (a) and $T_h/T_c=3.0$ (b).

Two features can be read off directly from Fig.~\ref{fig:eta_ratio}. First, the peak efficiency, optimized over $\Delta$, stays in a narrow band at about $0.72$--$0.75$ of the Carnot efficiency for all temperature ratios studied. Second, the qualitative dependence on $J$---the sharp rise, the broad decay, and the final exit from the engine mode---is preserved for all ratios.

Beyond these direct observations, a closer inspection of the figure shows that the role of the anisotropy $\Delta$ changes markedly with the temperature bias. At the small ratio $T_h/T_c=1.5$ the engine mode survives only for $\Delta\gtrsim1$: for $\Delta=0$ and $0.5$ no value of $J$ yields positive net work, while for $\Delta\geq1$ operation is confined to a narrow window centered near $J\approx3$, with a maximum efficiency of about $0.24$ at $\Delta=2.0$. A finite anisotropy is therefore a necessary resource at small temperature bias. At the large ratio $T_h/T_c=3.0$ the engine operates over a wide range of $J$ for all $\Delta$, and near the efficiency peak $\eta$ decreases monotonically with $\Delta$ (from about $0.48$ at $\Delta=0$ to $0.38$ at $\Delta=2.0$), reflecting the degradation of the effective compression ratio as $\Delta$ increases; at $T_h/T_c=2.0$ (main text) the dependence is non-monotonic, interpolating between these two limits.

\section*{CRediT authorship contribution statement}
\textbf{Meiru Li:} Validation, Software, Writing -- review \& editing.
\textbf{Maimaitiyiming Tusun:} Conceptualization, Methodology, Formal analysis, Investigation, Writing -- original draft.
\textbf{Fang Zhao:} Data curation, Formal analysis, Investigation, Writing -- review.
\textbf{Hasiyatihan Abudoula:} Data curation, Investigation.
\textbf{Tongcheng Wei:} Investigation, Writing -- original draft.

\section*{Declaration of competing interest}
The authors declare that they have no known competing financial interests or personal relationships that could have appeared to influence the work reported in this paper.

\section*{Data availability}
All data generated or analysed during this study are included in this published article. All numerical and symbolic computations, as well as the generation of all figures, were performed using open-source Python libraries (NumPy, SciPy, Matplotlib) only. No proprietary commercial software was used. The complete Python source code is available from the corresponding author upon reasonable request.

\end{document}